\documentclass[prb,twocolumn]{revtex4-2}

    \usepackage{graphicx}\usepackage{hyperref}
    \usepackage[left=1.5cm, right=1.5cm, top=2cm, bottom=2cm]{geometry}
\begin{document}

\title{Competing effects of Si and Mg doping on native point defects in yttrium aluminum garnet}

\author{M.\,Y.\,Vovsianiker$^1$, L.\,Yu.\,Kravchenko$^1$, and D.\,V.\,Fil$^{1,2}$}

\email{dmitriifil@gmail.com}

\affiliation{
$^1$Institute for Single Crystals, National Academy of Sciences of Ukraine,
60 Nauky Avenue, Kharkiv 61072, Ukraine\\
$^2$V.N. Karazin Kharkiv National University, 4 Svobody Square, Kharkiv 61022,
Ukraine}
\begin{abstract}
We investigate how Si and Mg doping affect native point defects in yttrium aluminum garnet (YAG) by combining first-principles calculations with a thermodynamic model of defect equilibria. Formation energies and equilibrium concentrations of native defects, dopants, and defect complexes are obtained across the full range of oxygen chemical potentials relevant to ceramic sintering. Donor (Si) and acceptor (Mg) dopants are found to control native defect populations efficiently but in opposite directions: Si substantially raises cation-vacancy concentrations while suppressing oxygen vacancies, whereas Mg produces the reverse trend. Depending on dopant concentration and oxygen chemical potential, native vacancy concentrations vary by up to thirteen orders of magnitude. The concentration of isolated cation vacancies depends nonmonotonically on Si content, a consequence of Si-vacancy complex formation. Under Si-Mg co-doping, the two dopants compensate each other efficiently, giving rise to pronounced minima in both cation- and oxygen-vacancy concentrations when Si and Mg are present in equal amounts. These findings provide a microscopic basis for the pronounced sensitivity of YAG sintering behavior to the Si/Mg ratio, and show that balanced donor-acceptor co-doping is an effective strategy for controlling native defect populations in YAG.

\end{abstract}

\maketitle

    \section{Introduction}

Yttrium aluminum garnet (YAG), doped with rare-earth or transition-metal ions, is widely used as a laser host material. YAG laser crystals can be obtained by traditional single-crystal growth or by ceramic technology, with YAG ceramics capable of reaching optical and mechanical characteristics nearly equal to those of single crystals \cite{1}. Ceramic technology offers a number of practical advantages, including the ability to fabricate elements with high dopant concentrations and to produce multilayer samples with different dopant concentrations in each layer.
At the same time, the optical quality achievable in transparent YAG ceramics remains strongly dependent on residual porosity, secondary-phase inclusions, and grain-boundary characteristics -- features whose extent and distribution are governed largely by the chosen sintering parameters \cite{2,3}.

Densification and microstructural control in YAG ceramics are commonly achieved through the use of sintering additives \cite{20}. Silicon-based additives -- typically introduced as tetraethyl orthosilicate (TEOS) or SiO$_2$ -- are among the most widely used in YAG sintering \cite{4,5,6,7,8,8b,8c}. Si is thought to aid densification primarily by generating cation vacancies that enhance mass transport, though this comes at the cost of accelerated grain growth, which lengthens the diffusion path along grain boundaries from the interior to the surface and reduces the efficiency of pore removal, ultimately limiting optical quality.

MgO acts in a complementary fashion: rather than enhancing cation diffusion, it primarily suppresses grain growth and recrystallization, thereby allowing densification to proceed more completely before pores become trapped within grains. Recent experimental studies \cite{8a,9,10} have demonstrated the effectiveness of MgO as a sintering aid for inhibiting grain growth in YAG ceramics. Reference \cite{9} showed that a 0.03 wt.\% MgO addition promoted fully dense, transparent YAG ceramics with transmittance close to the theoretical value for an ideal YAG crystal. Reference \cite{10} found that 0.01 wt.\% MgO doping inhibited recrystallization and greatly reduced pore content, with ceramics doped at 0.03-0.06 wt.\% MgO exhibiting an almost pore-free microstructure; at higher MgO content (0.06-0.1 wt.\%), however, the optical quality degraded again, with a detectable reappearance of residual pores and secondary phases.

Because Si and Mg additives act through largely independent mechanisms -- one enhancing cation diffusion, the other limiting grain growth — they are frequently combined \cite{11,12,13,14,15,16,17,18,19}, with SiO$_2$ eliminating porosity and MgO limiting grain size to jointly improve densification; the MgO content used is typically kept lower than that of SiO$_2$. Strikingly, the resulting microstructure depends sensitively on the relative proportions of the two additives: Reference \cite{19} found that YAG ceramics sintered with SiO$_2$ + MgO at equal atomic concentrations of Si and Mg contained an extremely high density of residual pores, scattering incident light strongly enough, making the ceramics nearly opaque. In contrast, ceramics sintered with an excess of either SiO$_2$ or MgO showed markedly higher optical quality, particularly those enriched in silicon, which contained very few pores.

Comparative work has further shown that Mg-doped ceramics tend to have finer, more uniform grains than their Si-doped counterparts, but at the expense of increased UV absorption linked to charge compensation around the Mg$^{2+}$ ion; Si$^{4+}$ doping, in contrast, tends to preserve or improve transmittance in the visible range \cite{Zhang23}. The optimal Si$^{4+}$/Mg$^{2+}$ ratio has also been found to depend on the specific surface area of the starting powders \cite{Wang23}, consistent with the two dopants occupying distinct lattice sites and being compensated through different mechanisms.

Because point-defect populations underlie diffusion, sintering behavior, and defect clustering, resolving how Si and Mg alter these populations is central to understanding their processing roles. Earlier atomistic studies based on empirical pair potentials proposed that Si substitution is compensated by cation vacancies, whereas Mg substituting for cations is compensated either by oxygen vacancies or by Mg interstitials \cite{111}. This studies also suggested that divalent Mg preferentially associates with oxygen vacancies to form neutral defect clusters. Subsequent DFT calculations reproduced these results and additionally predicted a thermodynamic preference for Si-Mg pairing under co-doping \cite{112}, but did not establish the equilibrium charged-defect concentrations under realistic sintering conditions, nor the significance of dopant-native-defect complexes in setting those concentrations.

Equilibrium defect concentrations depend on both the electron and oxygen chemical potentials, the latter set by the oxygen partial pressure during sintering. Most prior first-principles studies restrict their analysis to the O-rich and O-poor bounds of YAG stability; while informative as limiting cases, these bounds do not capture the continuous variation of the oxygen chemical potential encountered under realistic sintering atmospheres -- conditions under which defect equilibria can deviate considerably from the two extremes. The electron chemical potential, meanwhile, is not freely adjustable but is fixed self-consistently by the charge-neutrality condition once the oxygen chemical potential is specified.

Here, we combine first-principles calculations with a thermodynamic model of charged defects to obtain equilibrium concentrations of native defects, dopants, and their complexes across the full range of oxygen chemical potentials relevant to YAG sintering, with particular emphasis on how the competing donor (Si) and acceptor (Mg) dopants compete to shape native vacancy populations. The results offer a microscopic account of the contrasting behavior of Si and Mg additives and show how their combined use can be tuned to control native point-defect populations in YAG.

\section{Computational Method}
\subsection{DFT Calculations}

The crystal structure of YAG ($\mathrm{Y}_3\mathrm{Al}_5\mathrm{O}_{12}$) belongs to the space group $Ia\bar{3}d$. The cubic unit cell contains eight formula units (160 atoms). Y atoms occupy the dodecahedral 24(c) Wyckoff positions, while Al atoms occupy the octahedral 16(a) and tetrahedral 24(d) Wyckoff positions. O atoms occupy the 96(h) Wyckoff positions.

The formation energies of native defects, impurities, and defect complexes were calculated using density functional theory (DFT) as implemented in the SIESTA code~\cite{Soler}. The calculations employed the PBE exchange--correlation functional, norm-conserving PseudoDojo pseudopotentials~\cite{Dojo}, and a double-zeta polarized atomic-orbital basis set. Lattice parameters and atomic positions were optimized until the residual forces were below 0.005 eV/\AA\ and the residual stress components were smaller than 0.05 GPa. Defect formation energies were evaluated using the fully optimized 160-atom unit cell. A single isolated defect or defect complex was introduced into the unit cell, followed by relaxation of the atomic positions. The total energy of the relaxed defective structure was then determined.

Since the PBE functional significantly underestimates the band gap of YAG, the positions of the valence-band maximum ($E_\mathrm{VBM}$) and conduction-band minimum ($E_\mathrm{CBM}$) were determined independently using the Quantum ESPRESSO package~\cite{QE} with a modified PBE0 hybrid functional (mixing parameter $\alpha=0.32$). The calculated band gap of 7.7 eV is in good agreement with the experimental fundamental band gap of approximately 8 eV~\cite{fgap1,fgap2,fgap3}.

SIESTA employs a localized atomic-orbital basis, whereas Quantum ESPRESSO uses a plane-wave basis. Consequently, the absolute PBE eigenvalues of the VBM and CBM are not identical in the two codes. However, the hybrid-functional corrections to the band-edge positions are much less sensitive to the basis representation and are found to be very similar in independent hybrid-functional calculations performed with VASP~\cite{hp1,hp2}. Therefore, rather than using the absolute band-edge positions obtained with Quantum ESPRESSO, we transferred only the hybrid-functional corrections to the SIESTA calculations. The SIESTA band structure was corrected by applying rigid shifts of $-2.1$ eV to the VBM and $+1.15$ eV to the CBM, yielding the corrected band gap of 7.7 eV. These corrected band-edge positions were subsequently used in the calculation of charged-defect formation energies.

\subsection{Calculation of the Madelung correction}

The formation energies of point defects were evaluated using the Zhang--Northrup formalism~\cite{Zhang1991}:

\begin{eqnarray}
E^\mathrm{f}=E_{\mathrm{def}}-E_\mathrm{perf}
+\sum_{A}(r_{A}-a_{A})\mu_{A}
+q\mu_{e}
+E_{c},
\label{eq1}
\end{eqnarray}
where $E_{\mathrm{def}}$ and $E_\mathrm{perf}$ are the total energies of the defective and perfect supercells, respectively. The quantities $\mu_A$ and $\mu_e$ denote the chemical potentials of atomic species $A$ and electrons, $r_A$ and $a_A$ are the numbers of atoms removed from and added to the supercell to form the defect, $q$ is the defect charge state, and $E_c$ is the finite-size correction introduced to eliminate electrostatic interactions between periodically repeated charged defects.

The correction $E_c$ was determined using the extrapolation scheme proposed in Refs.~\cite{h09,h10} and implemented for Al$_2$O$_3$ and YAG in Refs.~\cite{my2} and \cite{vovs}, respectively. Formation energies without finite-size correction, $E^{\mathrm{f}(0)}$, were calculated for the cubic unit cell and for $1\times1\times2$ and $1\times1\times3$ supercells. These values were fitted to
\begin{eqnarray}
E^{\mathrm{f}(0)}(n\times m\times p)=k_1+k_2v_M(n\times m\times p),
\label{eq2}
\end{eqnarray}
where $v_M(n\times m\times p)$ is the Madelung constant of a periodic lattice of unit charges (one charge per a $n\times m\times p$ supercell) in a neutralizing background. The fitting parameter $k_1$ gives the defect formation energy free of periodic-image interactions.  The correction applied to the $1\times1\times1$ cell is therefore
$
E_c=k_1-E^{\mathrm{f}(0)}(1\times1\times1).
$

For all representative charged defects considered, $E^{\mathrm{f}(0)}$ exhibits an approximately linear dependence on $v_M$ \cite{vovs}, indicating that the leading  finite-size error is well described by the $q^2/L$ term. Unlike the Makov--Payne~\cite{mp} and Freysoldt--Neugebauer--Van de Walle (FNV)~\cite{fnv} approaches, higher-order corrections, including quadrupole and potential-alignment terms~\cite{komsa}, were neglected. The observed linear scaling indicates that these contributions are small for the supercell sizes employed in the present work.

For practical purposes, the finite-size correction can be approximated by
$
E_c \approx \alpha q^2,
$
where $\alpha=0.13$ eV. This value agrees well with the electrostatic estimate
$
\alpha \approx {1.42 e^2}/({\varepsilon a})\approx0.15~\mathrm{eV},
$
evaluated for the YAG lattice constant, $a$, and the static dielectric constant $\varepsilon=11.7$.

\subsection{Construction of phase diagrams}

The chemical potentials entering Eq.~(\ref{eq1}) were determined from the thermodynamic equilibrium between YAG and competing phases. To ensure internal consistency, the phase diagram was constructed using the total energies of all relevant compounds calculated within the same DFT framework, employing the same exchange--correlation functional, pseudopotentials, basis sets, and computational parameters as in the defect formation energy calculations.

Table~\ref{t1} summarizes the DFT-calculated formation enthalpies of the relevant compounds and the DFT-calculated molecular binding energy of O$_2$, together with the available calorimetric data at $T=0$ or $T=298$~K. The DFT formation enthalpies are obtained from the DFT energy of the O$_2$ molecule and those of the corresponding elemental and compound phases. The DFT O$_2$ binding energy is defined as the difference the DFT energies of the O$_2$ molecule and two isolated O atoms.

The DFT calculations performed using the SIESTA code overestimate the binding energy of the oxygen molecule and consequently predict less negative formation enthalpies for the oxides than the experimental values. This is consistent with previous plane-wave DFT calculations \cite{wang06,dop3}.

To improve the agreement between the calculated and experimental formation enthalpies, we introduce an empirical correction to the energy of the O$_2$ molecule, $\Delta E_{{\rm O}_2}^{\rm emp}=0.94$~eV. With this correction, the calculated formation enthalpies of YAG, Al$_2$O$_3$, and the Mg-based compounds become nearly identical to the experimental values. The discrepancies for the remaining phases are substantially reduced, as well. 

We note that using an O$_2$ energy derived from the experimental molecular binding energy and the calculated energy of an isolated oxygen atom, instead of the DFT energy of O$_2$ with the empirically fitted correction $\Delta E_{{\rm O}_2}^{\rm emp}$, would result in an overcorrection, yielding formation enthalpies that are more negative than the calorimetric values.

\begin{table}

   \caption{Calculated ($H_{\rm DFT}$) and calorimetric ($H_{\rm cal}$) formation enthalpies of the oxides and the corresponding differences, $\Delta H=H_{\rm DFT}-H_{\rm cal}$, per oxygen atom. The last row gives the calculated and experimental binding energies of the O$_2$ molecule.} \label{t1}
  \begin{tabular}{|l|c|c|c|}
    \hline
       Compound&\multicolumn{2}{|c|}{Formation enthalpy$^*$ (eV)}& $\Delta H$    \\ \cline{2-3}&$H_{\rm DFT}$&$H_{\rm cal}$&per oxygen atom\\
        \hline
    Al$_2$O$_3$& -15.79 
 & -17.24  \cite{nist} ($T=0$ K)& 0.48\\
   Y$_2$O$_3$& -18.94 
 & -20.03 \cite{morss} ($T=298$ K)&0.37\\
   Y$_3$Al$_5$O$_{12}$& -68.69
 & -74.38$^{**}$ \cite{lin} ($T=298$ K)&0.47\\ 
  YAlO$_3$ & -17.63
 & -18.94 \cite{navr} ($T=298$ K)&0.44\\
 SiO$_2$& -8.25
 & -9.39 \cite{nist} ($T=0$ K)&0.57\\
 MgO& -5.73
 & -6.19  \cite{nist} ($T=0$ K)&0.46\\
 MgAl$_2$O$_4$& -21.82
 & -23.66 \cite{nist} ($T=0$ K)&0.46\\
  Y$_2$SiO$_5$& -27.99
 & -29.96 \cite{liang} ($T=298$ K)&0.40\\
    O$_2$&-6.91&-5.12 \cite{nist} ($T=0$ K)&-0.90\\
        \hline  
  \end{tabular}
$^{*}$ For O$_2$, the values correspond to its molecular binding energy.

$^{**}$ Calculated using the enthalpy of formation of YAG from YAlO$_3$ and Al$_2$O$_3$ \cite{lin}.

\end{table}

\begin{figure}
\begin{center}
\includegraphics[width=8cm]{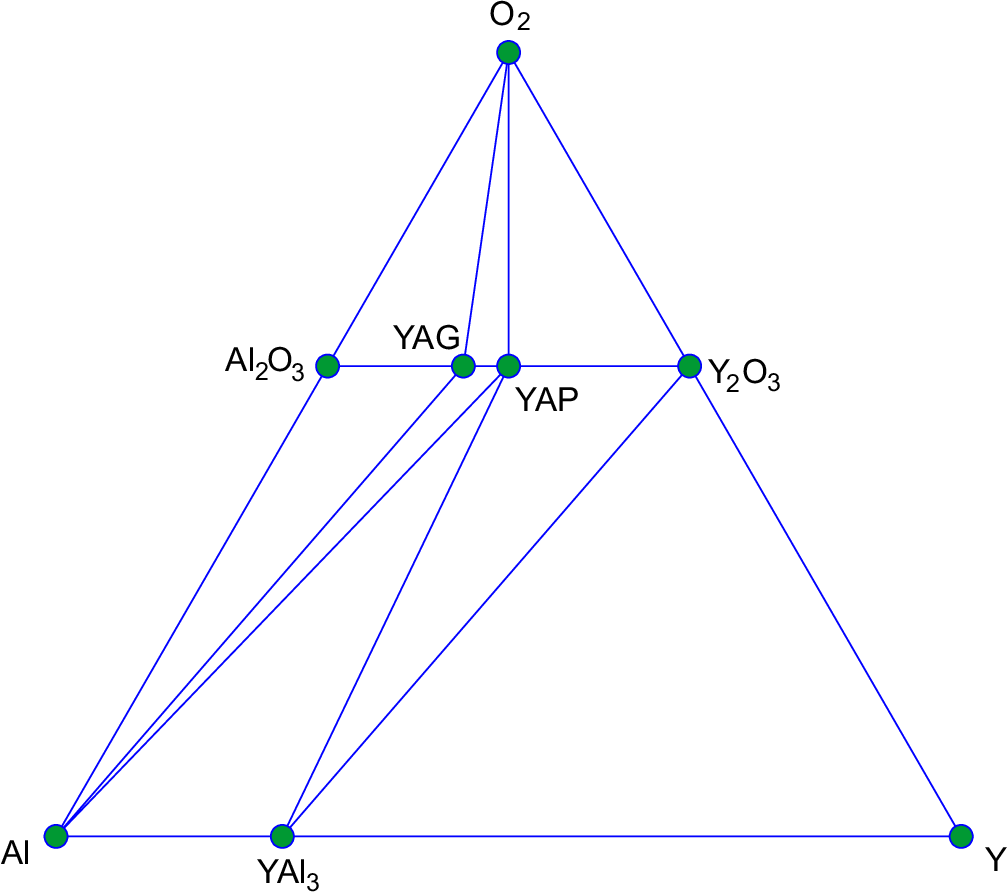
}\end{center}
\caption{Phase diagram of the $\mathrm{Y}$--$\mathrm{Al}$--$\mathrm{O}$ system. The abbreviations YAG and YAP stand for Y$_3$Al$_5$O$_{12}$ garnet and YAlO$_3$ perovskite, respectively.} \label{f1}
\end{figure}

\begin{figure}
\begin{center}
\includegraphics[width=8cm]{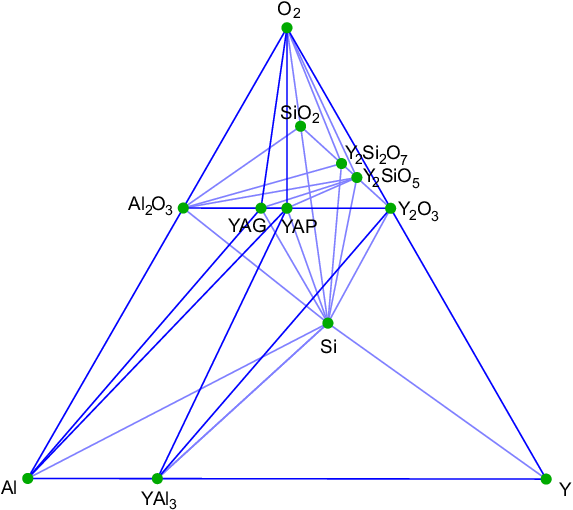}\end{center}
\caption{Phase diagram of the $\mathrm{Y}$--$\mathrm{Al}$--$\mathrm{O}$--$\mathrm{Si}$ system.} \label{f2}
\end{figure}

\begin{figure}
\begin{center}
\includegraphics[width=8cm]{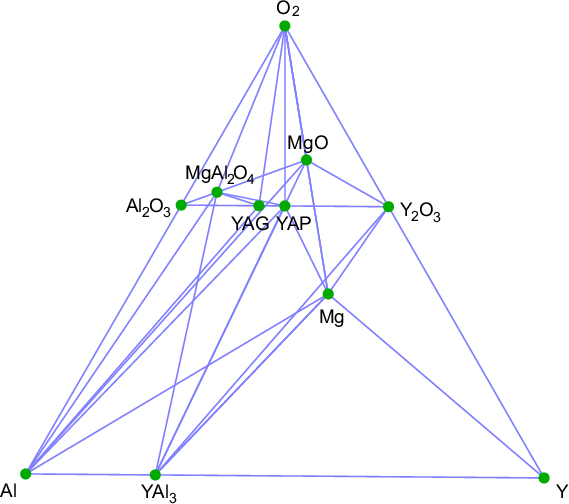}\end{center}
\caption{Phase diagram of the Y--Al--O--Mg system.} \label{f3}
\end{figure}

The calculated phase diagram of the Y--Al--O system is shown in Fig.~\ref{f1}. Since our primary objective is to determine the thermodynamic limits relevant for defect calculations in YAG, only phases that can coexist in equilibrium with YAG are included. The corresponding phase diagrams of the Y--Al--O--Si and Y--Al--O--Mg systems are presented in Figs.~\ref{f2} and \ref{f3}, respectively.

Figure~\ref{f1} reveals four three-phase equilibrium points involving YAG. Two of them, (YAG, Al$_2$O$_3$, Al) and (YAG, YAlO$_3$ , Al), define the oxygen-poor (reducing) limit of YAG stability, whereas the other two, (YAG, Al$_2$O$_3$, O$_2$) and (YAG, YAlO$_3$, O$_2$), correspond to the oxygen-rich (oxidizing) limit. These equilibrium points define the boundaries of the accessible range of oxygen chemical potential commonly considered in first-principles calculations of defect formation energies. They can also be classified according to the cation-rich limit. The equilibria involving Al$_2$O$_3$ correspond to Al-rich (Y-poor) conditions, whereas those involving YAlO$_3$ (YAP) correspond to Y-rich (Al-poor) conditions.

The same four limiting equilibrium points are preserved in the quaternary Y--Al--O--Mg system. In this case, the fourth equilibrium phase is MgAl$_2$O$_4$ (spinel), while the limiting oxygen and cation chemical potentials remain unchanged. The Y--Al--O--Si system exhibits more complex behavior. In addition to the four equilibrium points found in the ternary system, two additional four-phase equilibrium states appear in which YAG coexists with elemental Si, Y$_2$SiO$_5$, and Al$_2$O$_3$ or YAP. These competing phases do not alter the oxygen-rich and oxygen-poor stability limits of YAG.

The phase diagrams shown in Figs.~\ref{f1}--\ref{f3} correspond to $T=0$ K because they are constructed directly from DFT total energies. At finite temperatures, the appropriate thermodynamic potential is the Gibbs free energy,
\begin{eqnarray}
G(T,P)=E_{\mathrm{DFT}}+H(T,P)-H(0,P)-TS(T,P), \label{eq4a}
\end{eqnarray}
where $H(T,P)-H(0,P)$ and $S(T,P)$ represent the thermal enthalpy and entropy contributions, respectively.

Among all species considered in the present work, gaseous oxygen exhibits by far the largest temperature dependence owing to its large translational and rotational entropy. Furthermore, its Gibbs free energy depends much more strongly on pressure than that of condensed phases. The temperature- and pressure-dependent correction to the oxygen chemical potential for an ideal gas is given by  \cite{frey}
\begin{eqnarray}
\Delta\mu_{\mathrm{O}_2}(T,P)=k_\mathrm{B}T\left[\ln\!\!\left(\frac{P V_q}{k_\mathrm{B}T}\right)+\ln\!\!\left(\frac{\sigma_{\mathrm{O}_2} \Theta_{\mathrm{O}_2}}{T}\right)\right]  +\mu_{\mathrm{O}_2}^{\mathrm{vib}}, \label{eq4}
\end{eqnarray}
where $P$ is oxygen partial pressure, $V_q=(2\pi\hbar^2/m_{\mathrm{O}_2} k_\mathrm{B}T)^{3/2}$ is the translational quantum volume, $\Theta_{\mathrm{O}_2}=2.07$ K is the rotational temperature, $\sigma_{\mathrm{O}_2}=2$ is the molecular symmetry factor, and $\mu_{\mathrm{O}_2}^{\mathrm{vib}}$ is the vibrational free-energy contribution.

The temperature dependence of the Gibbs energies of formation is therefore dominated by the contribution of the gaseous oxygen phase. If temperature corrections to the Gibbs energies of condensed phases are included, they should be treated consistently with the corresponding temperature corrections to the defect-supercell and perfect-supercell energies in Eq.~(\ref{eq1}). The latter can be evaluated using the approach of Refs.~\cite{my2}. But for the charge compensation mechanisms considered in this paper the latter two types of corrections compensate each other and practically do not charge the concentration of relevant defect species.  Consequently, these corrections are not taken into account in the present work, and only the temperature dependence of the oxygen chemical potential is retained. 

In the following, we set the sintering temperature to $T=2023$ K ($t=1750^\circ$C) for all calculations.
At $P=P_0=0.1$ MPa (normal pressure), Eq.~(\ref{eq4}) yields
\begin{equation}
\Delta\mu_{\mathrm{O}_2}(2023\ \mathrm{K},P_0)=-4.67\ \mathrm{eV}+\mu_{\mathrm{O}_2}^{\mathrm{vib}}. \label{eq4b}
\end{equation}
Alternatively, using the tabulated enthalpy and entropy of oxygen from the NIST thermochemical database \cite{nist}, we obtain
\begin{equation}
\Delta\mu_{\mathrm{O}_2}(2023\ \mathrm{K},P_0)=-4.76\ \mathrm{eV}. \label{eq4c}
\end{equation}
Comparison of Eqs.~(\ref{eq4b}) and (\ref{eq4c}) shows that the vibrational contribution at 2023 K is only about $-0.09$ eV. In the following calculations, we therefore use the value given by Eq.~(\ref{eq4c}), which includes the vibrational contribution.

The calculated chemical potentials of Y and Al for equilibrium between YAG and either Al$_2$O$_3$ or YAlO$_3$ differ by less than 0.02 eV. Such a small difference has a negligible effect on the calculated defect formation energies. Accordingly, we consider only the chemical potentials corresponding to equilibrium between YAG and Al$_2$O$_3$.

 Although the phase diagrams shown in Figs.~\ref{f1}--\ref{f3} identify several characteristic equilibrium points, the oxygen chemical potential is not restricted to these discrete values. Instead, it varies continuously with the oxygen partial pressure and can be treated as an independent thermodynamic variable within the stability region of YAG. The lower bound of the accessible range, $\mu_{\mathrm O}^{\min}$, is determined by the stability limit of Al$_2$O$_3$, whereas the upper bound is taken as
$$
\mu_{\mathrm O}^\mathrm{max}=\frac{1}{2}\mu_{\mathrm{O}_2}(2023~\mathrm K,P_0),$$ which corresponds to molecular oxygen at the standard pressure. The quantity $\mu_{\mathrm O}^\mathrm{max}$ includes the empirical correction $\Delta E_{{\rm O}_2}^{\rm emp}/2$ discussed above. Throughout this work, $\mu_{\mathrm O}^{\max}$ is used as the reference value for the oxygen chemical potential.

The chemical potential of Mg is determined from the equilibrium condition with MgAl$_2$O$_4$ (spinel), using its DFT formation energy. For Si, the chemical potential is bounded by the stability conditions imposed by both elemental Si and Y$_2$SiO$_5$. Accordingly, at each value of the oxygen chemical potential, the thermodynamically allowed Si chemical potential is taken as the lower of the values imposed by these two competing phases.

By continuously varying the oxygen chemical potential within the interval defined above, we describe the entire range of thermodynamically accessible sintering conditions, from strongly reducing environments approaching equilibrium with metallic Al to oxidizing conditions corresponding to sintering in air.

\subsection{Equilibrium Defect Concentrations}
\label{s2d}

The equilibrium concentrations of point defects were determined by minimizing the Gibbs free energy of the crystal within the dilute-defect approximation. Assuming that defects do not interact with one another except through interactions within defect complexes, the Gibbs free energy can be written as \cite{frey}
\begin{equation}
F=F_0+\sum_i E_i n_i-k_{\rm B}T\ln W,
\label{eq5}
\end{equation}
where $F_0$ is the Gibbs free energy of the perfect crystal, $E_i$ is the formation energy of the $i$th defect species, including the temperature-dependent contribution from the oxygen chemical potential, $n_i$ is the number of defects of species $i$, and $W$ is the configurational degeneracy. In the dilute-defect approximation,
\begin{equation}
W=\prod_i\frac{N_i!}{(N_i-n_i)!n_i!},
\end{equation}
where $N_i$ denotes the number of lattice sites available for defects of species $i$.

Minimization of Eq.~(\ref{eq5}) with respect to $n_i$ yields the equilibrium occupation of the corresponding lattice sites by defect species $i$,
\begin{equation}
\tilde n_i=\frac{n_i}{N_i}
=\exp\left(-\frac{E_i}{k_{\rm B}T}\right),
\label{eq6}
\end{equation}
which is valid in the dilute-defect limit ($\tilde n_i\ll1$), where $T$ is the sintering temperature. The corresponding concentration of defect species $i$ is
$$c_i=N_i^{\rm cell}\tilde n_i/{\Omega_0},$$ where $\Omega_0$ is the unit-cell volume and $N_i^{\rm cell}$ is the number of symmetry-equivalent sites per unit cell.

The electron chemical potential, $\mu_e$,  is determined from the charge-neutrality condition \cite{frey}
\begin{equation}
\sum_i q_i c_i+c_h-c_e=0,
\label{eq7}
\end{equation}
where $c_h$ and $c_e$ are the concentrations of holes and electrons, respectively. Assuming nondegenerate carrier statistics, they are given by
\begin{equation}
c_h={C_h(T)}
\exp\left[-\frac{\mu_e-E_{\rm VBM}}{k_{\rm B}T}\right],
\label{eq8}
\end{equation}
and
\begin{equation}
c_e={C_e(T)}
\exp\left[-\frac{E_{\rm CBM}-\mu_e}{k_{\rm B}T}\right],
\label{eq9}
\end{equation}
where $E_{\rm VBM}$ and $E_{\rm CBM}$ are the valence-band maximum and conduction-band minimum, respectively. The prefactors $C_h(T)$ and $C_e(T)$ are evaluated directly from the DFT electronic structure as
\begin{equation}
C_h(T)=
\frac{2}{(2\pi)^3}
\sum_{i\in VB}\int_{BZ}d\mathbf{k}
\exp\left[
-\frac{E_{\rm VBM}-E_{i\mathbf{k}}}
{k_{\rm B}T}
\right]
,
\label{eq10}
\end{equation}

\begin{equation}
C_e(T)=
\frac{2}{(2\pi)^3}
\sum_{i\in CB}\int_{BZ}d\mathbf{k}
\exp\left[
-\frac{E_{i\mathbf{k}}-E_{\rm CBM}}
{k_{\rm B}T}
\right]
,
\label{eq11}
\end{equation}
where $E_{i{\bf k}}$ are the DFT eigenvalues, $i$ is the energy-band index, and the sums in Eqs.~(\ref{eq10}) and (\ref{eq11}) run over all bands below and above the fundamental band gap, respectively. The concentration of free carriers can also be evaluated by integrating the calculated electronic density of states \cite{frey}. In the case of isotropic parabolic bands, the free-carrier concentrations can be expressed in terms of the electron and hole effective masses \cite{frey}.

Equations~(\ref{eq6})--(\ref{eq11}) are solved self-consistently to determine the electron chemical potential, $\mu_e$. The equilibrium concentrations of all defect species are then evaluated using the resulting $\mu_e$.

The defect concentrations are calculated at the sintering temperature $T=2023$ K. During subsequent cooling, the defects are assumed to become immobile. Charge exchange between defects and free carriers reduces the concentration of free carriers \cite{vovs}, while the concentrations of distinct defect species remain frozen.

The solubility of dopants may differ from their nominal concentration introduced into the crystal during synthesis. To model experimentally a prescribed dopant concentration, the free-energy minimization is performed under an additional concentration constraint using the method of Lagrange multipliers.

For Si-doped YAG, the corresponding thermodynamic potential is
\begin{equation}
G=F_0+\sum_i E_i n_i-k_{\rm B}T\ln W-\lambda_\mathrm{Si}\left(\sum_i k_i^{\rm Si} n_i-n_{\mathrm{Si,tot}}\right),
\label{eq12}
\end{equation}
where $k_i^{\rm Si}$ is the number of Si atoms contained in defect species $i$, $n_{\rm Si,tot}$ is the total number of Si atoms in the crystal, and $\lambda_{\rm Si}$ is the Lagrange multiplier.

Minimization of Eq.~(\ref{eq12}) yields
\begin{equation}
\tilde n_i=
\exp\left[
-\frac{E_i-\lambda_{\rm Si}k_i^{\rm Si}}
{k_{\rm B}T}
\right].
\label{eq13}
\end{equation}

The Lagrange multiplier is determined simultaneously with the electron chemical potential from the charge-neutrality condition, Eq.~(\ref{eq7}), and the constraint
\begin{equation}
\sum_i k_i^{\rm Si}n_i=
n_{\rm Si,tot}.
\label{eq14}
\end{equation}

The Lagrange multiplier may be interpreted as an effective chemical potential of Si required to maintain the prescribed total dopant concentration. Consequently, the  concentrations of defect species, Eq. (\ref{eq13}),  are independent of the Si chemical potential obtained from the phase equilibrium of the Y--Al--O--Si system. Instead, they are determined by the total Si concentration together with the chemical potentials of Y, Al, and O defining the thermodynamic stability region of YAG.

The same formalism is readily extended to Mg-doped and Si--Mg co-doped YAG systems by introducing analogous concentration constraints for Mg or simultaneously for both dopants.

  \section{Defect Energetics and Concentrations}

\subsection{Defect formation energies}
 
\begin{figure}
\begin{center}
\includegraphics[width=8cm]{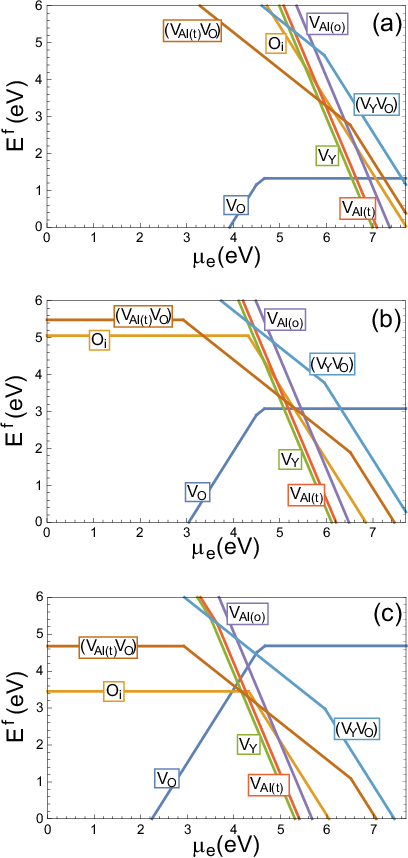}\end{center}
\caption{Formation energies of native point defects in YAG under reducing (a), intermediate (b), and oxidizing (c) conditions. Different slopes correspond to different charge states.} \label{f4}
\end{figure}

Figure~\ref{f4} shows the calculated formation energies of native point defects in YAG as  functions of the electron chemical potential for three representative oxygen chemical potentials. The symbols Al(t) and Al(o) denote Al atoms occupying tetrahedral and octahedral crystallographic sites, respectively. Throughout this paper, point defects are described using the Kröger--Vink notation. The electron chemical potential is varied across the band gap from the valence-band maximum, $E_{\rm VBM}$, to the conduction-band minimum, $E_{\rm CBM}$. The bandgap edge correction is applied
to the formation energy plots. For clarity, only the thermodynamically stable charge state of each defect is shown at each electron chemical potential. The three panels correspond to reducing conditions ($\mu_{\rm O}=\mu_{\rm O}^{\rm min}$), intermediate conditions ($P_{\rm O_2}=10^{-3}$ Pa, $\mu_{\rm O}=\mu_{\rm O}^{\rm max}-1.6$ eV), and oxidizing conditions ($P_{\rm O_2}=P_0$, $\mu_{\rm O}=\mu_{\rm O}^{\rm max}$).

Figure~\ref{f4} indicates that the dominant charged native defects in undoped YAG are negatively charged cation vacancies and positively charged oxygen vacancies, which together satisfy the charge-neutrality condition. Among the isolated cation vacancies, Y vacancies have the lowest formation energies, approximately 0.3 eV lower than those of tetrahedral Al vacancies, whereas octahedral Al vacancies have formation energies about 1.1 eV higher than those of Y vacancies. 

The strong association of  Al an Y vacancies with oxygen vacancies  leads to the formation of stable  complexes. These complexes may dominate the equilibrium population of negatively charged native defects. 
 The binding energies of vacancy-vacancy complexes are listed in Table~\ref{t2}. The $(V_{\rm Al(t)}^{'''}V_{\rm O}^{\bullet\bullet})^{'}$ complex has the most negative binding energy, $-4.13$ eV. The binding energies of other vacancy-vacancy complexes are smaller and their energies are higher. In particular, the formation energies of $(V_{\rm Al(o)}^{'''}V_{\rm O}^{\bullet\bullet})^{'}$ and $(V_{\rm Al(o)}^{'''}V_{\rm O}^{\bullet})^{''}$ (not shown in Fig.~\ref{f4}) are approximately 2.0 and 2.6 eV higher, respectively, than those of the corresponding complexes involving tetrahedral Al vacancies.

\begin{table}

   \caption{Binding energies of vacancy-vacancy complexes.   } \label{t2}
  \begin{tabular}{|l|c|}
    \hline
    Complex &  Binding energy (eV)  \\ \hline   $(V_\mathrm{Al(t)}^{\prime\prime\prime}V_\mathrm{O}^{\bullet\bullet})^\prime$ & -4.13 \\   $(V_\mathrm{Y}^{\prime\prime\prime}V_\mathrm{O}^{\bullet\bullet})^\prime$& -2.51 \\   $(V_\mathrm{Al(o)}^{\prime\prime\prime}V_\mathrm{O}^{\bullet\bullet})^\prime$& -2.92 \\  $(V_\mathrm{Al(t)}^{\prime\prime\prime}V_\mathrm{O}^{\bullet})^{\prime\prime}$ & -2.11 \\
$(V_\mathrm{Y}^{\prime\prime\prime}V_\mathrm{O}^{\bullet})^{\prime\prime}$ & -1.05  \\
$(V_\mathrm{Al(o)}^{\prime\prime\prime}V_\mathrm{O}^{\bullet})^{\prime\prime}$& -0.37 \\
    \hline
  \end{tabular}

\end{table}

The formation energies of most native defects are relatively high, resulting in low equilibrium concentrations under typical sintering conditions. The only notable exception is the oxygen vacancy under strongly reducing conditions, where the formation energy of $V_{\rm O}^\times$ becomes sufficiently low to yield an appreciable equilibrium concentration. Antisite defects, $\mathrm{Y}_{\mathrm{Al}}$ and $\mathrm{Al}_{\mathrm{Y}}$, are also known to occur in significant concentrations (see, for instance, \cite{vovs}). However, because their concentrations are expected to be only weakly affected by Si or Mg doping, they are not considered further in the present work.

\begin{figure}
\begin{center}
\includegraphics[width=8cm]{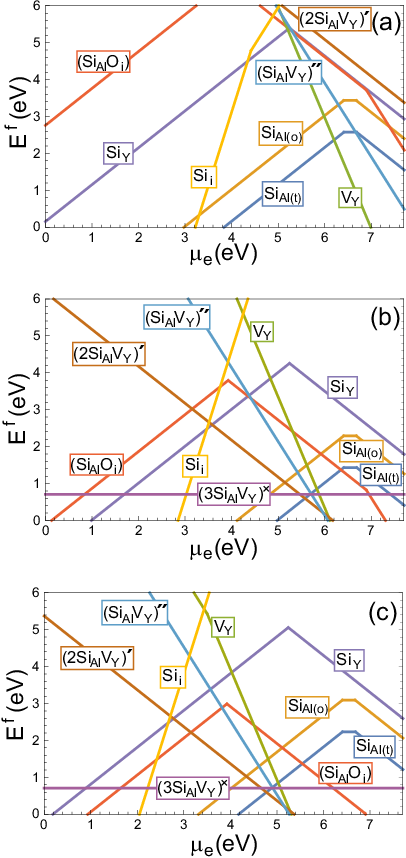}\end{center}
\caption{Formation energies of Si-related point defects in Si-doped YAG under reducing (a), intermediate (b), and oxidizing (c) conditions.} \label{f5}
\end{figure}

Figure~\ref{f5} shows the calculated formation energies of Si-related point defects in YAG as functions of the electron chemical potential. For comparison, the formation energy of  Y vacancies is also included.

The calculations indicate that Si is incorporated into YAG predominantly by substituting  Al atoms, forming the $\mathrm{Si}_{\mathrm{Al(t)}}^{\bullet}$ defects. The formation energies of $\mathrm{Si}_{\mathrm{Al(o)}}^{\bullet}$ and $\mathrm{Si}_{\mathrm{Y}}^{\bullet}$ are higher by approximately 0.86 and 4.0 eV, respectively, while interstitial Si has an even larger formation energy. 
Consequently, thermodynamic equilibrium strongly favors occupation of tetrahedral Al sites, and more than 99\% of the incorporated Si atoms are predicted to occur as $\mathrm{Si}_{\mathrm{Al(t)}}^{\bullet}$ defects.

Because substitutional Si acts as a singly charged donor, its incorporation must be accompanied by negatively charged defects to maintain charge neutrality. This role is played primarily by cation vacancies. Accordingly, Si doping increases the concentrations of $V_{\mathrm Y}^{'''}$ and $V_{\mathrm{Al}}^{'''}$, as discussed in the following subsection. In addition to isolated vacancies, substitutional Si exhibits a pronounced tendency to associate with cation vacancies, forming stable defect complexes such as $(\mathrm{Si}_{\mathrm{Al}}^{\bullet}V_{\mathrm Y}^{'''})^{''}$, $(2\mathrm{Si}_{\mathrm{Al}}^{\bullet}V_{\mathrm Y}^{'''})^{'}$, $(\mathrm{Si}_{\mathrm{Al}}^{\bullet}V_{\mathrm{Al}}^{'''})^{''}$, and $(2\mathrm{Si}_{\mathrm{Al}}^{\bullet}V_{\mathrm{Al}}^{'''})^{'}$.  For each complex, all possible configurations were considered, and Table~\ref{t3} reports the binding energy of the configuration with the most negative value.

\begin{table}

   \caption{Binding energies of complexes formed by Si$_{\rm Al}$ substitutional defects and cation vacancies.   } \label{t3}
  \begin{tabular}{|l|c|}
    \hline
    Complex &  Binding energy (eV)  \\ \hline   $({\rm Si}_\mathrm{Al(t)}^{\bullet}V_\mathrm{Al(t)}^{\prime\prime\prime})^{\prime\prime}$ & -0.80 \\   $({\rm Si}_\mathrm{Al(t)}^{\bullet}V_\mathrm{Al(o)}^{\prime\prime\prime})^{\prime\prime}$ & -1.13 \\   $({\rm Si}_\mathrm{Al(t)}^{\bullet}V_\mathrm{Y}^{\prime\prime\prime})^{\prime\prime}$ & -1.23 \\  $({\rm Si}_\mathrm{Al(o)}^{\bullet}V_\mathrm{Al(t)}^{\prime\prime\prime})^{\prime\prime}$  & -1.28 \\$({\rm Si}_\mathrm{Al(o)}^{\bullet}V_\mathrm{Y}^{\prime\prime\prime})^{\prime\prime}$  & -1.15 \\
$(2{\rm Si}_\mathrm{Al(t)}^{\bullet}V_\mathrm{Al(t)}^{\prime\prime\prime})^{\prime}$   & -1.36  \\
$(2{\rm Si}_\mathrm{Al(t)}^{\bullet}V_\mathrm{Al(o)}^{\prime\prime\prime})^{\prime}$ & -2.06 \\$(2{\rm Si}_\mathrm{Al(t)}^{\bullet}V_\mathrm{Y}^{\prime\prime\prime})^{\prime}$ & -2.21 \\$({\rm Si}_\mathrm{Al(t)}^{\bullet}{\rm Si}_\mathrm{Al(o)}^{\bullet}V_\mathrm{Al(t)}^{\prime\prime\prime})^{\prime}$ & -1.70 \\$({\rm Si}_\mathrm{Al(t)}^{\bullet}{\rm Si}_\mathrm{Al(o)}^{\bullet}V_\mathrm{Y}^{\prime\prime\prime})^{\prime}$ & -1.88 \\$(3{\rm Si}_\mathrm{Al(t)}^{\bullet}V_\mathrm{Y}^{\prime\prime\prime})^{\times}$ & -2.70 \\
    \hline
  \end{tabular}
\end{table}

The calculated binding energies indicate that complexes involving Y vacancies are generally more stable than the corresponding complexes containing Al vacancies. Therefore, only Y-vacancy complexes are shown in Fig.~\ref{f5}. Under oxidizing and intermediate conditions, these complexes exhibit formation energies lower than those of the corresponding isolated defects. As a result, their formation contributes significantly to the equilibrium defect population and to the charge balance of the crystal. The formation of Si--vacancy complexes shifts the electron chemical potential and affects the concentrations of all charged defects.

\begin{figure}
\begin{center}
\includegraphics[width=8cm]{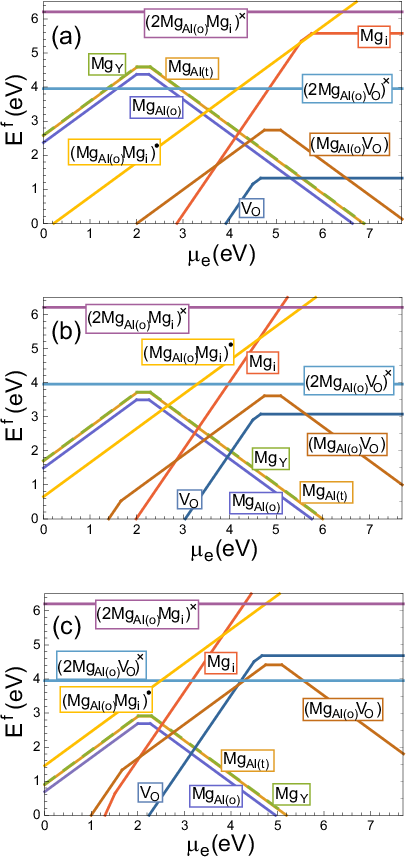}\end{center}
\caption{Formation energies of Mg-related point defects in Mg-doped YAG under reducing (a), intermediate (b), and oxidizing (c) conditions.}\label{f6}
\end{figure}

Figure~\ref{f6} shows the calculated formation energies of Mg-related point defects in Mg-doped YAG as functions of the electron chemical potential. The formation energy of oxygen vacancies is included as well, for comparison.

The calculations indicate that Mg is incorporated predominantly as a substitutional defect on both Al and Y sites. Among the substitutional configurations, $\mathrm{Mg}_{\mathrm{Al(o)}}^{\prime}$ has the lowest formation energy, although the energy differences between $\mathrm{Mg}_{\mathrm{Al(o)}}^{\prime}$, $\mathrm{Mg}_{\mathrm{Al(t)}}^{\prime}$, and $\mathrm{Mg}_{\mathrm{Y}}^{\prime}$ are quite small. In contrast, interstitial Mg defects and complexes of intersitial and substitutional Mg ions exhibit considerably higher formation energies and therefore have only a negligible influence on the defect population.

Because substitutional Mg acts as a singly charged acceptor, its incorporation is primarily compensated by positively charged oxygen vacancies. Consequently, Mg doping shifts the electron chemical potential in the direction opposite to that of Si doping, suppressing the concentrations of negatively charged cation vacancies while increasing the concentration of oxygen vacancies.

Substitutional Mg can associate with oxygen vacancies to form defect complexes. The binding energies of these complexes are summarized in Table~\ref{t4}. For clarity, Fig.~\ref{f6} shows only complexes involving the most energetically favorable substitutional defect, $\mathrm{Mg}_{\mathrm{Al(o)}}^{\prime}$. 

Although Mg-related complexes exhibit rather strong attractive interactions, their formation energies are relatively high. Consequently, their equilibrium concentrations remain lower than those of isolated substitutional Mg defects, and the complexes make a minor contribution to the overall charge neutrality of the defect ensemble.

\begin{table}

   \caption{Binding energies of complexes formed by Mg$_{\rm Al}$ and Mg$_{\rm Y}$ substitutional defects with oxygen vacancies and with intersitial Mg ions. } \label{t4}
  \begin{tabular}{|l|c|}
    \hline
    Complex &  Binding energy (eV)  \\ \hline   $({\rm Mg}_\mathrm{Al(t)}^{\prime}V_\mathrm{O}^{\bullet\bullet})^{\bullet}$ & -1.27 \\  $({\rm Mg}_\mathrm{Al(o)}^{\prime}V_\mathrm{O}^{\bullet\bullet})^{\bullet}$ & -0.81 \\   $({\rm Mg}_\mathrm{Y}^{\prime}V_\mathrm{O}^{\bullet\bullet})^{\bullet}$ & -1.22 \\  $(2{\rm Mg}_\mathrm{Al(t)}^{\prime}V_\mathrm{O}^{\bullet\bullet})^{\times}$   & -1.57 \\$(2{\rm Mg}_\mathrm{Al(o)}^{\prime}V_\mathrm{O}^{\bullet\bullet})^{\times}$   & -1.48    \\
$({\rm Mg}_\mathrm{Al(t)}^{\prime}{\rm Mg}_\mathrm{Al(o)}^{\prime}V_\mathrm{O}^{\bullet\bullet})^{\times}$   & -1.86  \\
$(2{\rm Mg}_\mathrm{Y}^{\prime}V_\mathrm{O}^{\bullet\bullet})^{\times}$  & -2.07 \\
$({\rm Mg}_\mathrm{Al(o)}^{\prime}{\rm Mg}_i^{\bullet\bullet})^\bullet$&-1.13\\ $(2{\rm Mg}_\mathrm{Al(o)}^{\prime}{\rm Mg}_i^{\bullet\bullet})^\times$&-1.22\\ 
    \hline
  \end{tabular}

\end{table}

 \subsection{Influence of Si and Mg doping on the concentration of anion and cation vacancies in YAG}

The equilibrium defect concentrations were calculated according to the procedure described in Sec.~\ref{s2d}. The calculations covered all isolated defects and defect complexes in all relevant charge states and crystallographically distinct configurations, including those omitted from Figs.~\ref{f4}-\ref{f6} and Tables~\ref{t3}, \ref{t4}.

\begin{figure}
\begin{center}
\includegraphics[width=8cm]{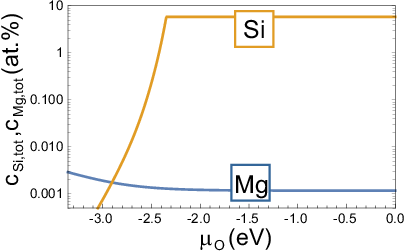}\end{center}
\caption{Equilibrium concentration of dopand atoms (in at.\%) in  YAG at $T=2023$ K. The reference value of the oxygen chemical potential is $\mu_{\rm O}^{\rm max} = 0.5 \mu_\mathrm{O_2}(\mathrm{2023\ K, 0.1\ MPa})$} \label{f7}
\end{figure}

Figure~\ref{f7} shows the total equilibrium concentrations of Si and Mg incorporated into YAG during sintering as functions of the oxygen chemical potential. The two curves correspond to Si and Mg doping, respectively; the case of simultaneous Si--Mg co-doping is discussed below. The concentration is given in atomic percent, relative to the total number of atoms in YAG, and can be converted to weight percent as $$ 1\ \mathrm{at. \%\ of\ Si}=7.02\ \mathrm{wt. \%\ of\ TEOS} =2.02\ \mathrm{wt. \%\ of\ SiO}_2,$$ $$ 
 1\ \mathrm{at. \%\ of\ Mg}=1.36\ \mathrm{wt. \%\ of\ MgO}.
$$

 Over most of the oxygen chemical potential range, the equilibrium dopant concentrations are nearly independent of $\mu_{\rm O}$. Under strongly reducing conditions, the equilibrium Si concentration decreases because of precipitation of elemental Si. The equilibrium Mg concentration, by contrast, increases under reducing conditions, because the combined formation energy of separate $\mathrm{Mg}^{\prime}$ and $V_{\rm O}^{\bullet}$ defects rises considerably toward the oxidizing limit, causing the dominant charge-compensating oxygen vacancies to change their charge state from $V_{\rm O}^{\bullet\bullet}$ to $V_{\rm O}^{\bullet}$.

The calculated equilibrium concentration of Si, $n_{\rm Si,tot}\approx 5$ at.\%, substantially exceeds the Si concentration typically achieved experimentally in YAG (about 0.15 at.\%) \cite{sun84}. Part of this discrepancy is expected because the predicted equilibrium concentration lies far outside the range of validity of the dilute-defect approximation: at the percent-level concentrations, the configurational entropy deviates substantially from its dilute-limit form, and interactions between defects beyond the isolated complexes considered here become significant. Both effects are expected to lower the true equilibrium concentration relative to the dilute-limit prediction.

The calculated Si concentration is, in addition, sensitive to the assumed secondary-phase equilibrium. The value $n_{\rm Si,tot}\approx 5$ at.\% quoted above corresponds to equilibrium between YAG and Al$_2$O$_3$, the case considered throughout this work. If instead equilibrium with YAP is assumed, the predicted total Si concentration is lower, $n_{\rm Si,tot}\approx 3$ at.\%. Both values remain roughly an order of magnitude above the experimentally reported solubility, indicating that the choice of competing phase alone does not resolve the discrepancy and that non-dilute corrections beyond the scope of the present defect model are additionally required to account for it.

It is worth noting that the non-dilute configurational corrections discussed above are expected to reduce the predicted equilibrium concentration relative to the dilute-limit estimate and therefore cannot explain a calculated concentration that is too low. For Mg, however, the calculated concentration is lower than the experimentally reported value. The calculated equilibrium concentration of Mg, $n_{\rm Mg,tot}\approx 1.0\cdot 10^{-3}$ at.\% ($1.2\cdot 10^{-3}$ under  equilibrium with YAP) is more than an order of magnitude lower than the Mg concentrations commonly reported for YAG (approximately 0.06 at.\%) \cite{10}. Since the corrections discussed above cannot raise a predicted concentration, this discrepancy must have a different origin: we attribute it to kinetic limitations on the precipitation of the competing spinel phase MgAl$_2$O$_4$.  Because nucleation and growth of a secondary phase during the finite sintering times used in ceramic processing (10--20 h in typical protocols \cite{Zhang23,Wang23}) can be slow relative to the incorporation of Mg into solid solution, the crystal may retain Mg concentrations exceeding the true thermodynamic solubility limit before phase separation occurs. This picture is consistent with energy-dispersive X-ray spectroscopy elemental mapping reported for MgO-doped Ce:YAG ceramics, which shows Mg-rich regions already present at the lowest investigated MgO addition (0.04 wt.\%), coexisting with regions where Mg, Al, and Y remain homogeneously distributed \cite{Zhang23} -- a pattern more indicative of incomplete progress toward a phase-separated equilibrium state than of a spatially uniform solubility limit. A directly comparable TEOS-doped series from the same study shows the converse tendency, with excess Si giving rise to Al$_2$O$_3$  segregation rather than a well-resolved secondary silicate phase \cite{Zhang23}, again suggesting that the experimentally accessible sintering conditions do not fully equilibrate with respect to secondary-phase formation.

Thus, although the calculations correctly reproduce the qualitative result that the solubility of Si in YAG exceeds that of Mg, they do not reproduce the absolute solubility limits quantitatively. Because absolute solubility limits are governed in part by growth kinetics that lie outside the scope of equilibrium defect thermodynamics, we do not attempt to reproduce them quantitatively. Instead, we vary the total Si and Mg concentrations within the experimentally relevant range (up to 0.1 at.\%) and use first-principles thermodynamics to determine the relative populations of the corresponding defect species at fixed total dopant concentrations, for which the dilute-defect approximation used here remains valid. In this approach, the relative concentrations of all defect species are determined entirely by the thermodynamics of YAG, see Sec. \ref{s2d}.

\begin{figure}
\begin{center}
\includegraphics[width=8cm]{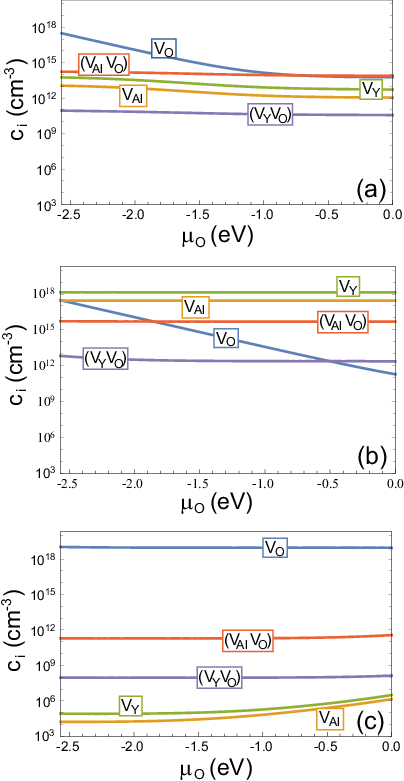}\end{center}
\caption{Concentrations of isolated vacancies and vacancy--vacancy complexes in undoped YAG (a), YAG doped with Si (b) and YAG doped with Mg (c) at a dopant concentration of 0.05 at.\%} \label{f8}
\end{figure}

To elucidate the influence of dopants on native defect concentrations, we calculated the latter as functions of the oxygen chemical potential for undoped YAG, Si-doped YAG, and Mg-doped YAG.  The oxygen chemical potential was bounded from below by  by the precipitation limit of elemental Si. The calculated concentrations are shown in Fig.~\ref{f8}. For the doped systems, the total dopant concentration was fixed at 0.05 at.\%. 

  Figure~\ref{f8} indicates that undoped and Si-doped YAG the concentrations of cation vacancies and their complexes with oxygen vacancies depend only weakly on the oxygen chemical potential, in both undoped and doped YAG. This weak dependence reflects the fact that the dominant charge-compensation mechanisms remain essentially unchanged over the wide range of $\mu_{\rm O}$. In contrast, the concentration of oxygen vacancies exhibits a pronounced dependence on the oxygen chemical potential because the dominant charge state of oxygen vacancies varies with $\mu_{\rm O}$. In undoped YAG, neutral oxygen vacancies, $V_{\rm O}^{\times}$, become the predominant oxygen-vacancy species over part of the reducing and intermediate oxygen chemical potential range. Si doping suppresses singly and doubly charged oxygen vacancies, making $V_{\rm O}^{\times}$ defects the dominant oxygen-vacancy species throughout almost the entire stability range. 
  
 In contrast, Mg doping stabilizes doubly positively charged oxygen vacancies, $V_{\rm O}^{\bullet\bullet}$, which remain the predominant oxygen-vacancy species over the entire range of oxygen chemical potentials shown in Fig. \ref{f8} and keep the total oxygen-vacancy concentration nearly constant. At the same time, the concentration of cation vacancies varies by an order of magnitude.
  
At the dopant concentration considered in Fig.~\ref{f8}, Si doping increases the concentrations of isolated Y and Al vacancies by up to five orders of magnitude, while reducing the concentration of oxygen vacancies  by approximately three orders of magnitude. The opposite effect is observed for Mg doping: the concentrations of cation vacancies decrease by up to eight orders of magnitude, whereas the concentration of oxygen vacancies  increases by as much as five orders of magnitude. These changes directly reflect the opposite roles of Si and Mg as donor and acceptor dopants, respectively.

Finally, Fig.~\ref{f8} demonstrates that vacancy--vacancy complexes make only a minor contribution to the equilibrium charge balance in both Si- and Mg-doped YAG. Although these complexes possess substantial binding energies, their  concentrations remain significantly lower than those of the dominant isolated charged defects.

\begin{figure}
\begin{center}
\includegraphics[width=8cm]{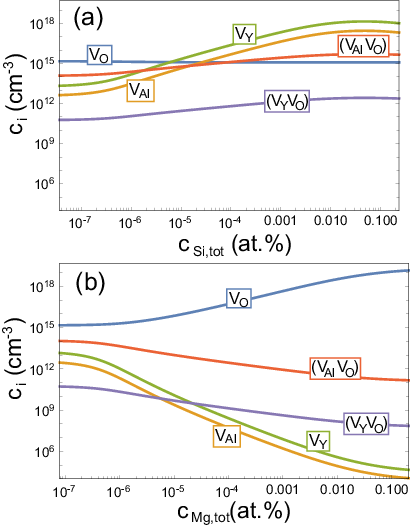}\end{center}
\caption{Concentrations of isolated vacancies and vacancy--vacancy complexes in Si-doped (a) and Mg-doped (b) YAG as functions of dopant concentration.} \label{f9}
\end{figure}

The dependence of the equilibrium concentrations of isolated vacancies and vacancy--vacancy complexes on the dopant concentration is shown in Fig.~\ref{f9}. The oxygen chemical potential corresponds to intermediate conditions ($P_{\rm O_2}=10^{-3}$ Pa, $\mu_{\rm O}=\mu_{\rm O}^{\rm max}-1.6$ eV).

Figure~\ref{f9}(a) shows that Si doping begins to modify the native defect population once the Si concentration exceeds approximately $10^{-6}$ at.\% ($\sim10^{15}$ cm$^{-3}$), which is comparable to the equilibrium concentration of cation vacancies in undoped YAG. Over the concentration range from about $10^{-6}$ to $5\times10^{-3}$ at.\% of Si, the concentrations of isolated cation vacancies increase nearly linearly with the Si content, reflecting charge compensation of substitutional $\mathrm{Si}_{\rm Al}^{\bullet}$ donors by negatively charged cation vacancies. At higher Si concentrations, however, the increase gradually slows as vacancy concentrations are increasingly affected by the formation of ternary $(2\mathrm{Si}_{\rm Al}V_{\rm Al(Y)})^\prime$ complexes. At a Si concentration of approximately $5\times10^{-2}$ at.\%, the concentration of isolated cation vacancies reaches a maximum and subsequently decreases as the ternary complexes become the thermodynamically preferred charge-compensating defects.

The behavior of Mg-doped YAG, shown in Fig.~\ref{f9}(b), is similar at low concentrations but differs at higher concentrations, where it saturates rather than passing through a maximum. The concentration of isolated oxygen vacancies increases approximately linearly with the Mg content over the range from $10^{-5}$ to $10^{-2}$ at.\%, where isolated $\mathrm{Mg}_{\rm Al(Y)}^{\prime}$ defects are predominantly compensated by oxygen vacancies. At higher Mg concentrations, however, this increase gradually slows, because an increasing fraction of substitutional Mg atoms becomes incorporated into $(\mathrm{Mg}_{\rm Al(Y)}V_{\rm O})$ complexes rather than remaining as isolated defects, eventually leveling off at a saturation value rather than decreasing as in the Si-doped case.

  \subsection{Co-doping with Si and Mg}
  
When YAG is co-doped with Si and Mg, positively charged substitutional Si defects and negatively charged substitutional Mg defects can associate to form electrically neutral $(\mathrm{Si}_{\rm Al}\mathrm{Mg}_{\rm Al})^{\times}$ and  $(\mathrm{Si}_{\rm Al}\mathrm{Mg}_{\rm Y})^{\times}$ complexes. Their calculated binding and formation energies  are summarized in Table~\ref{t5}. Although the binding energies are only moderate, the formation energies of some complexes are remarkably low under intermediate and oxidizing conditions. These low formation energies indicate that the incorporation of Si--Mg complexes into YAG is energetically favorable and that the co-doped system approaches the instability limit with respect to the formation of of phases with high dopant content.

\begin{table}
\begin{center}
   \caption{Binding energies and formation energies of $(\mathrm{Si}_{\mathrm{Al}}\mathrm{Mg}_{\mathrm{Al}})$ and $(\mathrm{Si}_{\mathrm{Al}}\mathrm{Mg}_{\mathrm{Y}})$ complexes. The formation energies are given for intermediate and oxidizing conditions. } \label{t5}
  \begin{tabular}{|l|c|c|}
    \hline
Complex &  Binding energy & Formarion energy   \\ \hline   $({\rm Mg}_\mathrm{Al(o)}^{\prime}{\rm Si}_\mathrm{Al(t)}^{\bullet})^{\times}$ & -0.39 eV& 0.39 eV\\  $({\rm Mg}_\mathrm{Al(t)}^{\prime}{\rm Si}_\mathrm{Al(t)}^{\bullet})^{\times}$ & -0.29 eV&0.73 eV\\    $({\rm Mg}_\mathrm{Y}^{\prime}{\rm Si}_\mathrm{Al(t)}^{\bullet})^{\times}$& -0.40 eV&0.63 eV\\   $({\rm Mg}_\mathrm{Al(o)}^{\prime}{\rm Si}_\mathrm{Al(o)}^{\bullet})^{\times}$  & -0.22 eV&1.43 eV\\ $({\rm Mg}_\mathrm{Al(t)}^{\prime}{\rm Si}_\mathrm{Al(o)}^{\bullet})^{\times}$   & -0.40 eV & 1.48 eV \\
 $({\rm Mg}_\mathrm{Y}^{\prime}{\rm Si}_\mathrm{o}^{\bullet})^{\times}$   & -0.48 & 1.42 eV\\

    \hline
  \end{tabular}
\end{center}
\end{table}

The calculated low formation energies are consistent with experimental observations demonstrating the formation of highly substituted single-phase YAG ceramics. In Ref.~\cite{wu22}, single-phase
$ \mathrm{Y}_3\mathrm{Mg}_x\mathrm{Al}_{5-2x}\mathrm{Si}_x\mathrm{O}_{12}$
ceramics were obtained for $x=0.5$, 1, and 1.5, whereas secondary phases appeared only at $x=2$. Similar results were reported in Ref.~\cite{du18} for
$ \mathrm{Y}_3\mathrm{Mg}_x\mathrm{Al}_{5-2x}\mathrm{Si}_x\mathrm{O}_{12}:\mathrm{Ce},$
where single-phase ceramics were successfully synthesized for Mg--Si co-doping levels up to at least $x=1$.

 At the dopant concentrations less than 0.1 at.\% the concentration of Mg--Si complexes remains low. Nevertheless, substitutional Mg and Si defects still compensate each other efficiently through the charge-neutrality condition, thereby substantially reducing the concentrations of native charged defects. Consequently, the effect of co-doping on the native defect population is governed by the excess dopant species, i.e., by the difference between the Si and Mg concentrations rather than by their absolute concentrations.

\begin{figure}
\begin{center}
\includegraphics[width=8cm]{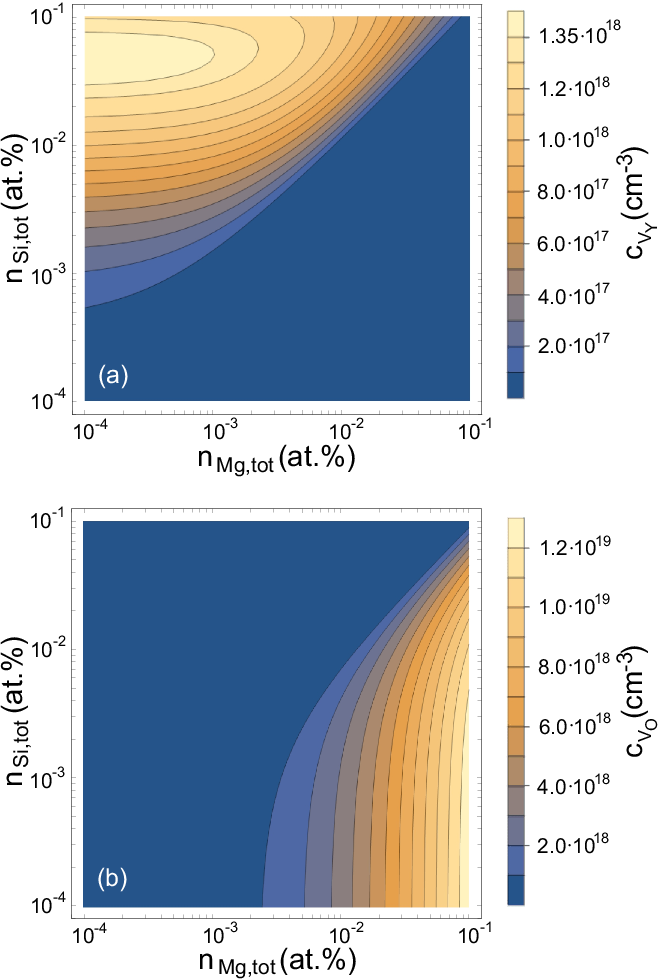}\end{center}
\caption{Concentrations of isolated yttrium vacancies (a) and oxygen vacancies (b) in YAG co-doped with Si and Mg as functions of the concentrations of dopant atoms during sintering under intermediate conditions ($\mu_\mathrm{O}=-1.6$ eV).}\label{f10}
\end{figure}

\begin{figure}
\begin{center}
\includegraphics[width=8cm]{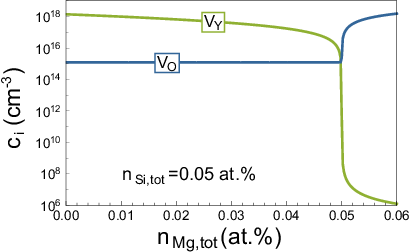}\end{center}
\caption{Concentrations of isolated yttrium and oxygen vacancies in YAG co-doped with Si and Mg at a constant Si concentration and varying Mg concentration during sintering under intermediate conditions ($\mu_\mathrm{O}=-1.6$ eV).} \label{f11}
\end{figure}

This behavior is illustrated in Figs.~\ref{f10} and \ref{f11}, which show the equilibrium concentrations of yttrium and oxygen vacancies as functions of the Si and Mg concentrations.

Figure~\ref{f10} demonstrates that the nonmonotonic dependence of the concentration of isolated yttrium vacancies as the function of the Si content persists under co-doping when the Si concentration exceeds that of Mg. The oxygen vacancy concentration remains a monotonic function of the Mg concentration, although its magnitude is reduced by simultaneous Si doping.

Figure~\ref{f11} illustrates the effect of Mg co-doping on YAG containing a fixed Si concentration. As the Mg concentration approaches the Si concentration, the concentration of isolated yttrium vacancies decreases abruptly by approximately ten orders of magnitude. Simultaneously, the oxygen-vacancy concentration increases, particularly under intermediate and oxidizing conditions. When the Si and Mg concentrations become equal, charge compensation occurs predominantly between the dopants rather than through native defects. Consequently, the concentrations of both cation and oxygen vacancies are reduced to values close to those in undoped YAG.

\section{Discussion}

The present calculations demonstrate that Si doping increases the equilibrium concentration of isolated cation vacancies by up to five orders of magnitude. Among these vacancies, yttrium vacancies are the dominant species, accounting for more than 80\% of the total cation-vacancy population over most of the stability range. Such a substantial increase in the concentration of Y vacancies unbound from Si$_{\rm Al}$ defects  is expected to enhance Y-ion diffusion during sintering and thereby contribute to the accelerated densification observed experimentally.

An important finding is that the concentration of isolated cation vacancies is a nonmonotonic function of the total Si concentration. The concentration reaches a maximum at a relatively low Si content, approximately 0.05~at.\%, after which it decreases because of the formation of Si--vacancy complexes. Consequently, further addition of Si does not further enhance cation transport, despite the increase in dopant concentration. Moreover, substantially lower Si concentrations can produce a significant enhancement of vacancy-mediated diffusion. For example, at a total Si concentration of only 0.001~at.\%, the concentration of isolated cation vacancies already reaches approximately 15\% of its maximum value.

The origin of this nonmonotonic behavior can be understood using a simple picture. For clarity, we consider only the dominant substitutional donor defect, $\mathrm{Si}_{\rm Al(t)}^{\bullet}$, isolated yttrium vacancies, and the dominant ternary defect complex $(2\mathrm{Si}_{\rm Al(t)}V_{\rm Y})^{\prime}$. Let us denote the concentrations of isolated substitutional Si defects and isolated yttrium vacancies by
$$x=n_{{\rm Si}_{\rm Al(t)}^\bullet}, \quad y=n_{V_{\rm Y}^{\prime\prime\prime}}.$$
According to the thermodynamic formalism developed in Ref.~\cite{my2}, the concentration of the ternary complex satisfies
$$n_{(2{\rm Si}_{\rm Al(t)}V_{\rm Y})^{\prime}}=\alpha x^2 y,$$
where the coefficient $\alpha\gg 1$ depends on the  binding energy and temperature. The charge-neutrality condition and the constraint on the total Si concentration then reduce to
\begin{eqnarray}\label{xy-eq}
 x-3y-\alpha x^2 y=0,\cr
 x+ 2\alpha x^2 y=n_{\rm Si, tot}.
\end{eqnarray}

In the dilute limit, $\alpha n_{\rm Si,tot}^2\ll 1$, complex formation is negligible ($\alpha x^2 y\to 0$ in both equation) Eqs.~(\ref{xy-eq}) reduce to $x=3y$ and $x=n_{\rm Si, tot}$, giving
$$n_{V_{\rm Y}^{\prime\prime\prime}}\approx n_{\rm Si, tot}/3,$$
that is, the concentration of isolated vacancies increases linearly with the Si concentration. In the opposite limit, $\alpha n_{\rm Si,tot}^2\gg1$, nearly all vacancies are bound into complexes, so Eqs.~(\ref{xy-eq}) can be approximated as $x=\alpha x^2 y$ and $x+2\alpha x^2 y=n_{\rm Si, tot}$, giving
$$n_{V_{\rm Y}^{\prime\prime\prime}}\approx  \frac{3}{\alpha n_{\rm Si, tot}}.$$
Thus, the exact solution of Eqs.~(\ref{xy-eq}) interpolates between these two limits and exhibits a maximum at intermediate Si concentrations. Physically, the nonmonotonic dependence results from the competition between two opposing processes: donor doping generates compensating cation vacancies, whereas at higher Si concentrations these vacancies are progressively consumed by the formation of stable Si-vacancy complexes.

Although this analysis has been formulated for Si-doped YAG, the underlying mechanism is quite general. A similar nonmonotonic dependence of the isolated-vacancy concentration can be expected in oxides where trivalent host cations are substituted by tetravalent dopants, provided the dopants form sufficiently stable dopant-vacancy complexes containing multiple dopant ions.

The behavior of Mg-doped YAG is qualitatively different. Charge compensation is provided primarily by doubly positively charged oxygen vacancies, while the dominant charged defect complexes contain only one substitutional Mg ion and one oxygen vacancy. Consequently, complex formation leads to saturation of $n_{V_{\rm O}^{\bullet\bullet}}$ at high Mg concentrations rather than to a pronounced maximum followed by a decrease, as observed for isolated cation vacancies in Si-doped YAG.

The calculations reveal several distinct regimes for Mg incorporation into Si-doped YAG. Mg can be accommodated at concentrations that exceed the total Si content, with incorporated Mg occupying Al and Y sites as negatively charged defects. At low relative Mg content, these acceptors are compensated mainly by positively charged Si$_{Al}^\bullet$ 
 defects, which partially, but not completely, reduces the concentration of isolated cation vacancies relative to Si-doped YAG alone. As the Mg concentration approaches that of Si, however, this partial compensation gives way to a much sharper effect: the concentration of isolated cation vacancies collapses by as much as twelve orders of magnitude, while the oxygen-vacancy concentration stays as low as it is in purely Si-doped material. Beyond this point, further increases in Mg content cause the oxygen-vacancy concentration to rise steadily.

These trends offer a microscopic picture of Mg's role during sintering. MgO is conventionally understood to suppress abnormal grain growth mainly by segregating to grain boundaries and slowing their migration via solute drag. The present results suggest that Si doping actually assists this process: by supplying an efficient charge-compensation channel for substitutional Mg, Si allows higher Mg concentrations to be incorporated into the lattice, potentially increasing the pool of Mg available for grain-boundary segregation and solute drag.

As long as Si remains in excess of Mg, its effect on cation-vacancy concentration is essentially unaffected by the presence of Mg, meaning Si continues to promote vacancy-mediated cation diffusion while Mg independently limits boundary migration -- the two dopants thus act in a complementary, largely non-interfering fashion, simultaneously aiding densification and restraining grain growth.

The picture changes qualitatively once Si and Mg concentrations become comparable. Here, donor-acceptor compensation between the two dopants is highly effective, driving down both cation- and oxygen-vacancy concentrations by several orders of magnitude. This suppresses Si's ability to enhance cation diffusion. 
Once Mg is present in clear excess over Si, oxygen vacancies rather than Si donors become the dominant compensating species for Mg acceptors, and the oxygen-vacancy concentration rises sharply -- potentially aiding oxygen transport during sintering. At sufficiently high Mg content, formation of the MgAl$_2$O$_4$ 
 spinel phase becomes thermodynamically favorable; the resulting second-phase particles at grain boundaries and triple junctions would be expected to add a Zener-pinning contribution to grain-growth suppression, on top of the point-defect effects discussed above.

\section{Conclusions}

Combining first-principles calculations with thermodynamic modeling of defect equilibria, we have examined how Si and Mg doping shape the native point-defect landscape of YAG under sintering-relevant conditions. Donor-type Si and acceptor-type Mg are found to act in opposition: Si raises the equilibrium cation-vacancy concentration while lowering that of oxygen vacancies, and Mg does the reverse. The magnitude of this effect is substantial -- native vacancy concentrations shift by up to thirteen orders of magnitude depending on dopant level and oxygen chemical potential.

One notable finding is that the isolated cation-vacancy concentration does not increase monotonically with Si content, but instead peaks at moderate Si levels before falling off at higher doping, as vacancies are increasingly bound into stable Si-vacancy complexes. This behavior is captured by a simple model, which shows that such nonmonotonic behavior should arise generally whenever a tetravalent dopant substituting for a trivalent lattice cation forms stable complexes with the vacancies that compensate it.

Under Si-Mg co-doping, mutual compensation between the donor and acceptor defects becomes highly efficient as their concentrations approach each other, producing sharp minima in both cation- and oxygen-vacancy populations. Near this compensation point, Si's enhancement of cation-vacancy-mediated diffusion is largely switched off while oxygen-vacancy levels remain suppressed; once Mg is added in excess, the oxygen-vacancy concentration rises again. Together, these results give a microscopic account of why YAG sintering behavior is so sensitive to the Si/Mg ratio.

Taken together, these findings indicate that Si and Mg cannot be treated as independent sintering additives, since their combined effect is set by the interplay between defect formation, charge compensation, and defect association, which depends mainly on their relative concentrations.

\section*{Acknowledgments}

We acknowledge the support of the National Academy of Sciences of Ukraine through Project No. 0125U000562, ``Investigation of the Fundamental Properties of Complex Structured Media and Nanosystems.'' M.Y.V. acknowledges support from the Scholarship of the National Academy of Sciences of Ukraine for Young Scientists.


\begin{thebibliography}{00}

 \bibitem{1} A. Ikesue, T. Kinoshita, Fabrication of high-performance polycrystalline
 Nd:YAG ceramics for solid state lasers,  J. Am. Ceram. Soc. \textbf{78},   1033 (1995). \url{https://doi.org/10.1111/j.1151-2916.1995.tb08433.x}

 \bibitem{2}  A. Ikesue, K. Yoshida, T. Yamamoto, I. Yamaga, Optical scattering centers in
polycrystalline Nd:YAG laser, J. Am. Ceram. Soc. \textbf{80}, 1517 (1997).  \url{https://doi.org/10.1111/j.1151-2916.1997.tb03011.x}


\bibitem{3}  A. Ikesue, K. Yoshida, Influence of pore volume on laser performance of Nd : YAG
ceramics, J. Mater. Sci. \textbf{34},  1189 (1999). \url{https://doi.org/10.1023/A:1004548620802}

\bibitem{20} J. Hostasa, F. Picelli, S. Hribalova, V. Necina, Sintering aids, their role and behaviour in the production of
transparent ceramics, Open Ceramics \textbf{7}, 100137  (2021). \url{https://doi.org/10.1016/j.oceram.2021.100137}

\bibitem{4}  A. Maitre, C. Salle, R. Boulesteix, J. -F. Baumard, Y. Rabinovitch, Effect of silica on the reactive
sintering of polycrystalline Nd:YAG ceramics, J. Am. Ceram. Soc. \textbf{91},
406  (2008).  \url{https://doi.org/10.1111/j.1551-2916.2007.02168.x}

\bibitem{5}  S. Kochawattana, A. Stevenson, S. H. Lee, M. Ramirez, V. Gopalan, J. Dumm, V.
K. Castillo, G. J. Quarles, G. L. Messing, Sintering and grain growth in SiO$_2$ doped
Nd:YAG, J. Eur. Ceram. Soc. \textbf{28},  1527 (2008). \url{https://doi.org/10.1016/j.jeurceramsoc.2007.12.006}

\bibitem{6}  R. Boulesteix, A. Maitre, J. -F. Baumard, C. Salle, Y. Rabinovitch, Mechanism of the
liquid-phase sintering for Nd:YAG ceramics, Opt. Mater. \textbf{31}, 711 (2009). \url{https://doi.org/10.1016/j.optmat.2008.04.005}

\bibitem{7}  R. Boulesteix, A. Maitre, J. -F. Baumard, Y. Rabinovitch, C. Salle, S. Weber, M. Kilo,
The effect of silica doping on neodymium diffusion in yttrium aluminum garnet
ceramics: implications for sintering mechanisms, J. Eur. Ceram. Soc. \textbf{29}, 
2517 (2009). \url{https://doi.org/10.1016/j.jeurceramsoc.2009.03.003}

 \bibitem{8} A.J. Stevenson, X. Li, M. A. Martinez, J. M. Anderson, D. L. Suchy, E. R. Kupp, E.
C. Dickey, K. T. Mueller, G. L. Messing, Effect of SiO$_2$ on densification and microstructure
development in Nd:YAG transparent ceramics, J. Am. Ceram. Soc. \textbf{94},
1380 (2011).  \url{https://doi.org/10.1111/j.1551-2916.2010.04260.x}

\bibitem{8c} J. Hostasa, L. Esposito, A. Piancastelli, Influence of Yb and Si content on the sintering and phase changes of Yb:YAG laser ceramics,  J. Eur. Ceram. Soc. \textbf{32}, 2949  (2012). \url{https://doi.org/10.1016/j.jeurceramsoc.2012.02.045}

\bibitem{8b} S. J. Pandey, M. Martinez, J. Hostasa, L. Esposito, M. Baudelet, and R. Gaume, Quantification of SiO$_2$ sintering additive in YAG transparent ceramics by laser-induced breakdown spectroscopy (LIBS), Opt. Mater. Express \textbf{7},  1666 (2017). \url{https://doi.org/10.1364/OME.7.001666}


 \bibitem{8a} Z. Lu, T. Lu, N. Wei, W. Zhang, B. Ma, J. Qi, Y. Guan, X. Chen, H. Wu, Y. Zhao, Effect
of air annealing on the color center in Yb: Y$_3$Al$_5$O$_{12}$ transparent ceramics with MgO
as sintering additive, Opt. Mater. \textbf{47},  292 (2015). \url{https://doi.org/10.1016/j.optmat.2015.05.043}


 \bibitem{9} T. Zhou, L. Zhang, S. Wei, L. Wang, H. Yang, Z. Fu, Q. Zhang, H. Chen, F. A. Selim, Q. Zhang,
MgO assisted densification
of highly transparent YAG ceramics and their microstructural evolution,
J. Eur. Ceram. Soc. \textbf{38},  687  (2018). \url{https://doi.org/10.1016/j.jeurceramsoc.2017.09.017}

 \bibitem{10}  I. Vorona, A. Balabanov, M. Dobrotvorska, R. Yavetskiy, O. Kryzhanovska,
L. Kravchenko, S. Parkhomenko, P. Mateychenko, V. Baumer, I. Matolinova, Effect
of MgO doping on the structure and optical properties of YAG transparent
ceramics, J. Eur. Ceram. Soc. \textbf{40},  861 (2020). \url{https://doi.org/10.1016/j.jeurceramsoc.2019.10.048}


 \bibitem{11} Y. Li, S. Zhou, H. Lin, X. Hou, W. Li, H. Teng, T. Jia, Fabrication of Nd:YAG
transparent ceramics with TEOS, MgO and compound additives as sintering aids,
J. Alloy. Compd. \textbf{502}, 225 (2010). \url{https://doi.org/10.1016/j.jallcom.2010.04.151}


 \bibitem{12}  H. Yang, X. Qin, J. Zhang, S. Wang, J. Ma, L. Wang, Q. Zhang, Fabrication of Nd:
YAG transparent ceramics with both TEOS and MgO additives, J. Alloy. Compd.
\textbf{509},  5274 (2011). \url{https://doi.org/10.1016/j.jallcom.2010.11.030}

 \bibitem{13}  W. Guo, Y. Cao, Q. Huang, J. Li, J. Huang, Z. Huang, F. Tang, Fabrication and laser
behaviors of Nd:YAG ceramic microchips, J. Eur. Ceram. Soc. \textbf{31},  
2241 (2011).  \url{https://doi.org/10.1016/j.jeurceramsoc.2011.05.020}

 \bibitem{14}  J. Li, F. Chen, W. Liu, W. Zhang, L. Wang, X. Ba, Y. Zhu, Y. Pan, J. Guo, Co-precipitation
synthesis route to yttrium aluminum garnet (YAG) transparent
ceramics, J. Eur. Ceram. Soc. \textbf{32}, 2971  (2012) . \url{https://doi.org/10.1016/j.jeurceramsoc.2012.02.040}

 \bibitem{15} H. Yang, X. Qin, J. Zhang, J. Ma, D. Tang, S. Wang, Q. Zhang, The effect of MgO
and SiO$_2$ codoping on the properties of Nd:YAG transparent ceramic, Opt. Mater.
\textbf{34},  940 (2012). \url{https://doi.org/10.1016/j.optmat.2011.05.029}

\bibitem{16}  L. Zhang, Y. Li, X. Li, H. Yang, X. Qiao, T. Zhou, Z. Wang, J. Zhang, D. Tang,
Characterization of spray granulated Nd: YAG particles for transparent ceramics, J.
Alloys Compd. \textbf{639} 244 (2015) . \url{https://doi.org/10.1016/j.jallcom.2015.02.229}

\bibitem{17}   R. Yin, J. Li, M. Dong, T. Xie, Y. Fu, W. Luo, L. Ge, H. Kou, Y. Pan, J. Guo,
Fabrication of Nd: YAG transparent ceramics by non-aqueous gelcasting and vacuum
sintering, J. Eur. Ceram. Soc.  \textbf{36}, 2543 (2016). \url{https://doi.org/10.1016/j.jeurceramsoc.2016.03.013}

\bibitem{18} F. Mohammadi, O. Mirzaee, M. Tajally, Influence of TEOS and MgO addition on slurry rheological, optical, and
microstructure properties of YAG transparent ceramic, Optical Materials \textbf{85},  174 (2018). \url{https://doi.org/10.1016/j.optmat.2018.08.047}


\bibitem{19} I.O. Vorona, R.P. Yavetskiy, S.V. Parkhomenko, A.G. Doroshenko, O.S. Kryzhanovska,
N.A. Safronova, A.D. Timoshenko, A.E. Balabanov, A.V. Tolmachev, V.N. Baumer, Effect of complex $\mathrm{Si}^{4+}+\mathrm{Mg}^{2+}$ additive on sintering and properties of
undoped YAG ceramics, J. Eur. Ceram. Soc.  \textbf{42}, 6104  (2022).  \url{https://doi.org/10.1016/j.jeurceramsoc.2022.05.017}

\bibitem{Zhang23} Q. Zhang, Y. Shi, D. Kosyanov, Y. Xiong, T. Wu, Z.-Z. Zhou, Q. Liu, J. Fang, J. Ni, H. He a, J. Yu, M. Niu, W. Liu,  Effect of extra added Mg$^{2+}$ and Si$^{4+}$ on the microstructure and luminescence properties of Ce:YAG ceramic phosphors for high power LED/LD lighting, Ceramics International \textbf{49}, 11311 (2023). \url{https://doi.org/10.1016/j.ceramint.2022.11.331}


\bibitem{Wang23} X. Wang, Y. Shi, C. Zhang, Z. Lu, R. Chen, J. Qi, T. Lu, Optimizing liquid phase promotion effect by modifying powder specific
surface area in Si$^{4+}$/Mg$^{2+}$ co-doped transparent Yb:YAG ceramics, Journal of the European Ceramic Society \textbf{43}, 6363 (2023). \url{https://doi.org/10.1016/j.jeurceramsoc.2023.06.022}




 \bibitem{111}  M. M. Kuklja, Defects in yttrium aluminium perovskite and
garnet crystals: atomistic study, J. Phys.: Condens. Matter \textbf{12}, 2953  (2000). \url{https://doi.org/10.1088/0953-8984/12/13/307}

\bibitem{112} S. Jiang, T. Lu, J. Chen, Ab initio study the effects of
Si and Mg dopants on point defects and Y diffusion in YAG, Computational Materials Science  \textbf{69},
 261 (2013).  \url{https://doi.org/10.1016/j.commatsci.2012.11.045}

\bibitem{Soler} J. M. Soler, E. Artacho, J. D. Gale, A. Garcia, J. Junquera, P. Ordejon, and
D. Sanchez-Portal,
The SIESTA method for ab initio order-N materials simulation, J. Phys.: Condens. Matter \textbf{14},  2745 (2002). \url{https://doi.org/10.1088/0953-8984/14/11/302}

\bibitem{Dojo} M. J. van Setten, M. Giantomassi, E. Bousquet, M. J. Verstraete, D. R. Hamann, X. Gonze, and G.-M. Rignanese,
“The PseudoDojo: Training and grading a 85 element optimized norm-conserving pseudopotential table,” Comput. Phys.Commun. 226, 39 (2018), \url{https://doi.org/10.1016/j.cpc.2018.01.012}

\bibitem{QE} P. Giannozzi et al., “Quantum ESPRESSO: A modular and open-source software project for quantum simulations of
materials,” J. Phys.: Condens. Matter 21, 395502 (2009), \url{https://doi.org/10.1088/0953-8984/21/39/395502}

\bibitem{fgap1}
M. Kučera, V. N. Kolobanov, V. V. Mikhailin, P. A. Orekhanov, and V. N. Makhov,
``Reflection spectra of some garnet and orthoferrite single crystals in vacuum ultraviolet,''
{Phys. Status Solidi B} \textbf{157}, 745 (1990),
\url{https://doi.org/10.1002/pssb.2221570227}

\bibitem{fgap2}
M. Kirm, A. Lushchik, Ch. Lushchik, and G. Zimmerer,
``Investigation of luminescent properties of pure and Ce$^{3+}$-doped Y$_3$Al$_5$O$_{12}$ crystals using VUV radiation,''
{Electrochem. Soc. Proc.} \textbf{99-40}, 113 (2000).

\bibitem{fgap3}
T. Zorenko, V. Gorbenko, S. Nizankovskiy, and Yu. Zorenko,
``Comparison of the luminescent properties of Y$_3$Al$_5$O$_{12}$:Pr crystals and films,''
{Acta Phys. Pol. A} \textbf{133}, 948 (2018),
\url{https://doi.org/10.12693/APhysPolA.133.948}

\bibitem{hp1}
C. Linderälv, D. Åberg, and P. Erhart,
``Luminescence quenching via deep defect states: A recombination pathway via oxygen vacancies in Ce-doped YAG,''
{Chem. Mater.} \textbf{33}, 73 (2021),
\url{https://doi.org/10.1021/acs.chemmater.0c02449}

\bibitem{hp2}
W. Lafargue-Dit-Hauret, M. Allix, B. Viana, S. Jobic, and C. Latouche,
``Computational analysis on native and extrinsic point defects in YAG using the metaGGA SCAN method,''
{Theor. Chem. Acc.} \textbf{141}, 58 (2022),
\url{https://doi.org/10.1007/s00214-022-02920-7}

\bibitem{Zhang1991}
S. B. Zhang and J. E. Northrup,
``Chemical potential dependence of defect formation energies in GaAs: Application to Ga self-diffusion,''
{Phys. Rev. Lett.} \textbf{67}, 2339 (1991),
\url{https://doi.org/10.1103/PhysRevLett.67.2339}

\bibitem{h09}N. D. M. Hine, K. Frensch, W. M. C. Foulkes, and M. W.
Finnis, Supercell size scaling of density functional theory for-
mation energies of charged defects, Phys. Rev. B  \textbf{79}, 024112 (2009). \url{https://doi.org/10.1103/PhysRevB.79.024112}

\bibitem{h10} N. D. M. Hine, P. D. Haynes, A. A. Mostofi, and M. C.
Payne, Linear-scaling density-functional simulations of charged
point defects in Al$_2$O$_3$ using hierarchical sparse matrix algebra,
J. Chem. Phys. \textbf{133},
114111  (2010). \url{https://doi.org/10.1063/1.3492379}

\bibitem{my2}  L. Yu. Kravchenko and D. V. Fil, Control of charge state of dopants in insulating crystals: Case study of Ti-doped sapphire, Phys. Rev. Research \textbf{2}, 023135 (2020). \url{https://doi.org/10.1103/PhysRevResearch.2.023135}

\bibitem{vovs}  M. Y. Vovsianiker,  D. V. Fil, Cooling-induced charge-state redistribution of
native defects in yttrium aluminum garnet, Functional Materials \textbf{33}, 328 (2026).    \url{https://dx.doi.org/10.15407/fm33.02.328}

\bibitem{mp}
G. Makov and M. C. Payne,
``Periodic boundary conditions in ab initio calculations,''
{Phys. Rev. B} \textbf{51}, 4014 (1995).
\url{https://doi.org/10.1103/PhysRevB.51.4014}

\bibitem{fnv}
C. Freysoldt, J. Neugebauer, and C. G. Van de Walle,
``Electrostatic interactions between charged defects in supercells,''
{Phys. Status Solidi B} \textbf{248}, 1067 (2011).
\url{https://doi.org/10.1002/pssb.201046289}

\bibitem{komsa}
H.-P. Komsa, T. T. Rantala, and A. Pasquarello,
``Finite-size supercell correction schemes for charged defect calculations,''
{Phys. Rev. B} \textbf{86}, 045112 (2012).
\url{https://doi.org/10.1103/PhysRevB.86.045112}




\bibitem{nist} M. W. Chase, Jr., NIST-JANAF Thermochemical Tables, 4th ed.
(American Chemical Society and American Institute of Physics,
New York, 1998).

\bibitem{morss} L. R. Morss, P. P. Day, C. Felinto, H. Brito, Standard molar enthalpies of formation of Y$_2$O$_3$, Ho$_2$O$_3$, and Er$_2$O$_3$ at the temperature 298.15 K, J. Chem. Thermodynamics \textbf{25}, 415 (1993). \url{https://doi.org/10.1006/jcht.1993.1045}

\bibitem{lin} I-C. Lin, A. Navrotsky, J.K. Richard Weber, P. C. Nordine, Thermodynamics of glass formation and metastable solidification of molten Y$_3$Al$_5$O$_{12}$, Journal of Non-Crystalline Solids \textbf{243}, 273 (1999). \url{https://doi.org/10.1016/S0022-3093(98)00814-X}

\bibitem{navr} O. Fabrichnaya, H. J. Seifert, R. Weiland, T. Ludwig, F. Aldinger and A, Navrotsky, Phase Equilibria and Thermodynamics in the Y$_2$O$_3$-Al$_2$O$_3$-SiO$_2$ System, Int. J. of Materials Research \textbf{92}, 1083 (2001).

\bibitem{liang} J.-J. Liang and A. Navrotsky,  Enthalpy of formation of rare-earth silicates Y$_2$SiO$_5$ and
Yb$_2$SiO$_5$ and N-containing silicate Y$_{10}$(SiO$_4$)$_6$N$_2$, J. Mater. Res. \textbf{14}, 1181 (1999). \url{https://doi.org/10.1557/JMR.1999.0158} 




\bibitem{wang06} L. Wang, T. Maxisch, and G. Ceder, Oxidation energies of transition metal oxides within the GGA+U framework, Phys. Rev. B \textbf{73}, 195107 (2006).  \url{https://doi.org/10.1103/PhysRevB.73.195107}


\bibitem{dop3} W. Jing, M. Liu, J. Wen, L. Ning, M. Yin, and C.-K. Duan, First-principles study of Ti-doped sapphire. I. Formation and optical transition properties of titanium pairs, Phys. Rev. B \textbf{104}, 165103  (2021). \url{https://doi.org/10.1103/PhysRevB.104.165103};
 W. Jing, M. Liu, J. Wen, L. Ning, M. Yin, and C.-K. Duan,  First-principles study of Ti-doped sapphire. II. Formation and reduction of complex defects, Phys. Rev. B \textbf{104}, 165104  (2021). \url{https://doi.org/10.1103/PhysRevB.104.165104}





\bibitem{frey}  C. Freysoldt, B. Grabowski, T. Hickel, J. Neugebauer,
 G. Kresse, A. Janotti, and C. G. Van de Walle,
 First-principles calculations for point defects in solids,
 Rev. Mod. Phys. \textbf{86}, 253  (2014). \url{https://doi.org/10.1103/RevModPhys.86.253}

\bibitem{sun84}  W.Y. Sun, X.T. Li, L.T. Ma, T.S. Yen (D.S. Yan), Solubility of Si in YAG, J. Solid State Chem.
\textbf{51},  315 (1984). \url{https://doi.org/10.1016/0022-4596(84)90348-7}


 \bibitem{wu22}  G. Wu, M. Ma, A. Li, K. Song,  A. Khesro, H. B. Bafrooei, E. Taheri-nassaj, S. Luo, F. Shi, S. Sun, and D. Wang,
Crystal structure and microwave dielectric properties of Mg$^{2+}$-Si$^{4+}$ co-modified yttrium aluminum garnet ceramics, Journal of Materials Science: Materials in Electronics  \textbf{33}, 4712 (2022) . \url{https://doi.org/10.1007/s10854-021-07661-0}

\bibitem{du18} Q. Du,   S. Feng,   H. Qin,   H. Hua, H. Ding,  L. Jia,   Z. Zhang,   J. Jiang and  H. Jiang, Massive red-shifting of Ce$^{3+}$ emission by Mg$^{2+}$ and Si$^{4+}$ doping of YAG:Ce transparent ceramic phosphors, J. Mater. Chem. C \textbf{6},  12200 (2018).    \url{https://doi.org/10.1039/C8TC03866J}


\end{thebibliography}
\end{document}